# (H, Li)Br and LiOH Solvation Bonding Dynamics: Molecular Nonbond Interactions and Solute Extraordinary Capabilities

Chang Q Sun[*]

**Abstract**

We resolved the O:H-O bond transition from the mode of ordinary water to its hydration in terms of its phonon stiffness (vibration frequency shift $\Delta\omega$), order of fluctuation (line width), and number fraction (phonon abundance, $f_x(C) = N_{hyd}/N_{total}$). The $f_x(C)$ follows $f_H(C) = 0$, $f_{Li}(C) \propto f_{OH}(C) \propto C$, and $f_{Br}(C) \propto 1 - \exp(-C/C_0)$ toward saturation with $C$ being the solute concentration. The invariant $df_x(C)/dC$ suggests that the solute forms a constantly sized hydration droplet without responding to interference of other ions because its hydrating $H_2O$ dipoles fully screen its electric field. However, the number inadequacy of the highly ordered hydration $H_2O$ dipoles partially screens the large Br⁻. The Br⁻ then interacts repulsively with other Br⁻ anions, which weakens its electric field and the $f_{Br}(C)$ approaches saturation at higher solute concentration. The consistency in the concentration trend of the $f_{LiBr}(C)$, the Jones–Dole viscosity $\eta(C)$, and the surface stress of LiBr solution clarifies their common origin of ionic polarization. The resultant energy of the solvent H-O exothermic elongation by O:⇔:O repulsion and the solute H-O endothermic contraction by bond-order-deficiency heats up the LiOH solution. An estimation of at least 0.15 eV (160% of the O:H cohesive energy of 0.1 eV) suggests that the H-O elongation is the main source heating up the solution, while the molecular motion, structure fluctuation, or even evaporation dissipates energy caped at 0.1 eV.

1. ## Introduction

The local physical-chemical properties of hydrogen bonds (O:H-O or HB with ":" being the electron lone pair of oxygen) in the hydration shells are quite different from those of the ordinary bulk water, which has attracted extensive research interest from various perspectives. Fine-resolution detection and consistently deep insight into the intra- and intermolecular interactions and their consequence on the solution properties have been an area of active study. Intensive pump-probe spectroscopic investigations have been conducted to pursue the mechanism behind molecular performance in the spatial and temporal domains. For instance, the sum frequency generation (SFG) spectroscopy resolves information on the molecular dipole orientation or the skin dielectrics, at the air-solution interface[1-2], while the ultrafast two-dimensional infrared absorption probes the solute or water molecular diffusion dynamics in terms of phonon lifetime and the viscosity of the solutions [3-4]. Various perspectives have been involved in the understanding of solvation dynamics, which include $H^+$ and $OH^-$ donation, electron lone pair ":" acceptance and donation, etc.[5-8]

---

[*] School EEE, Nanyang Technological University, Singapore 639798 (ecqsun@ntu.edu.sg)

Salt solutions demonstrate the Hofmeister effect [9-10] on regulating the solution surface stress and the solubility of proteins with possible mechanisms of structural maker and breaker[11-13], ionic specification[14], quantum dispersion[15], skin induction[16], quantum fluctuation[17], and solute-water interactions[18]. The performance of the excessive $H^+$ protons in acid solutions and the lone pairs in basic solutions has been approached in terms of "molecular structural diffusion"[19] with involvement of proton thermal hopping[20], proton tunneling[21] or fluctuating[22]. Mechanisms proposed by Grotthuss[19, 23], Eigen[24], Zundel[25] and their combinations [26-27] are currently popular. The excessive protons in acidic solutions, and as an inverse of protons, electron lone pairs in basic solutions, form an $H_9O_4^+$ complex in which an $H_3O^+$ core is strongly hydrogen-bonded to three $H_2O$ molecules and leave the lone pair of the $H_3O^+$ free[24], or form an $H_5O_2^+$ complex in which the proton is shuttling freely between two $H_2O$ molecules [25].

Increasing the chloride, bromide and iodide solute concentration shifts more the H-O stretching vibration mode to higher frequencies[28-29], while the $OH^-$ shifts the H-O mode to lower frequencies. These spectral changes are usually explained as the $Cl^-$, $Br^-$, and $I^-$ ions weakening of the surrounding H-bond (structure breakers) or the $OH^-$ strengthening of the H-bond (structure makers). The H-bond is often referred to the O:H nonbond that is the part of the hydrogen bond (O:H-O). An external electric field in the $10^9$ V/m order slows down water molecular motion and even crystallizes the system. The field generated by a $Na^+$ ion acts rather locally to reorient and even hydrolyze its neighboring water molecules according to MD computations[30]. However, HCl hydration fragments water clusters into smaller ones[31]. Studies of NaOH hydration in bulk water[32] and water clusters [33] revealed two processes of H-O spectral signal relaxations. One is the slow process on $200 \pm 50$ fs time scales and the other faster dynamics on 1–2 ps scales.

However, knowledge insufficiency about O:H-O bond cooperativity[34] has hindered largely the progress in understanding the solvation bonding dynamics, solute capabilities, and inter- and intramolecular interactions in the HBr, LiBr, and LiOH solutions as a collection of comparison. One has been hardly able to resolve the network O:H-O bond segmental cooperative relaxation induced by acid, base, salt solvation, or solute bond-order-deficiency (the bond order of a $HO^-$ with one H-O bond is lower than a $H_2O$ with two H-O bonds). It is yet to be known how the $H^+(H_3O^+)$, $OH^-$, $Li^+$ and $Br^-$ ions interact with water molecules and their neighboring solutes, and their impact on the performance of the solutions such as the surface stress, solution viscosity, solution temperature, and critical pressures and temperatures for phase transition [28, 35]. Our understanding is limited within the mode of proton motion or molecular drifting, but what is going on inside the molecules is much more fascinating[28, 36-37].

Aside from the interest in the O:H phonon relaxation[38], solute-solvent interaction length[39], structure making or breaking [40 41], solute motion dynamics and phonon relaxation lifetime, we turn to examine the intra- and intermolecular interactions and the solute capabilities of transiting the number and stiffness of the O:H-O bonds from the mode of ordinary water into the hydration shells. We showed recently that O:H-O bond polarization, H↔H interproton disruption, and O:⇔:O inter-lone-pair compression essentially govern the solute-solvent interactions in the respective H(Cl, Br, I)[36], Na(F, Cl, Br, I)[42] and (LI, Na, K)OH[37] solutions.

An extension of our efforts[36-37, 42] to the present HBr, LiBr, and LiOH solvation dynamics has led to consistent insight into the solute-solute interaction and the solute capabilities of mediating the solution properties. We found that only Br⁻-Br⁻ repulsion exists in the LiBr and HBr solutions without the presence of Li⁺-Li⁺ or Li⁺-Br⁻ interactions. The O:H-O polarization, H↔H anti-HB fragilization[36] and O:⇔:O super-HB[37] compression, and the bond-order-deficiency induced bond contraction[43] stem their relevant phonon relaxation, surface stress, Jones–Dole viscosity, and the solvation thermodynamics.

## 2. Principles
### 2.1 DPS Probed Bond Transition Information

The pump-probe time-dependent phonon spectroscopy probes the decay time of a known intramolecular vibration (H-O) phonon band intensity to derive the molecular motion dynamics through the solution viscosity and Stokes-Einstein relation for drift diffusivity[44], which is very much the same to optical fluorescent spectroscopy [45]. The signal lifetime is proportional in a way to the density and distribution of the defects and impurities. The impurity or defect states prevent the thermalization of the electrons transiting from the excited states to the ground for exciton (or electron-hole pair) recombination.

Comparatively, a Raman spectral peak features the Fourier transformation of all bonds vibrating in the same frequency from the real space, irrespective of their locations or orientations. The spectral peak shape shows the probability distribution, and the peak maximum corresponds to the bond stiffness of highest distribution. The peak area is the abundance that is proportional to the number of bonds being detected and the peak width to the structure order of the vibrating bonds[46-47]. The frequency shift $\Delta\omega_x$, in the first order approximation, features the stiffness of the segmental x stretching vibration as a function of its length $d_x$ and energy $E_x$[47],

$$\Delta\omega_x \propto \sqrt{E_x/\mu_x}/d_x \propto \sqrt{(k_x+k_c)/\mu_x}$$

(1)

The subscript $x$ = L denotes the O:H nonbond characterized by the stretching vibration frequency at ~200 cm$^{-1}$ and $x$ = H denotes the H-O bond phonon frequency of ~3200 cm$^{-1}$ in the bulk water. The $k_x$ and $k_C$ are the force constants or the second differentials of the intra/inter molecular interaction and O-O Coulomb coupling potentials. The $\Delta\omega_x$ also varies with the reduced mass $\mu_x$ of the specific $x$ oscillator. However, from the full-frequency Raman spectra, one could hardly be able to resolve the transition of bonds by solvation. Inclusion of high-order nonlinear interactions only offsets the peak position without adding any new features of vibrations[47-48].

A difference between the spectra collected after and before solvation (called DPS[49-50]) can resolve the transition of the phonon stiffness (frequency shift) and abundance (peak area) by solvation. The fraction coefficient, $f_x(C)$, being the integral of the DPS peak, represents the fraction of bonds, or the number of phonons transiting from water to the hydration states at a solute concentration $C$,

$$f_x(C) = \int_{\omega_m}^{\omega_M} \left[ \frac{I_{solution}(C,\omega)}{\int_{\omega_m}^{\omega_M} I_{solution}(C,\omega)d\omega} - \frac{I_{H_2O}(0,\omega)}{\int_{\omega_m}^{\omega_M} I_{H_2O}(0,\omega)d\omega} \right] d\omega.$$

The slope of the fraction coefficient, $df_x(C)/dC$, is proportional to the number of bonds per solute in the hydration shells, which characterizes the hydration shell size and its local electric field. The DPS distils only phonons transiting into their hydration states as a component presenting above the x-axis, which equals the abundance loss of the ordinary HBs as a valley below the axis in the DPS spectrum. This process removes the spectral areas commonly shared by the ordinary water and the high-order hydration shells. A hydration shell may contain one, two or more subshells, depending on the nature and size of the solute. The size and charge quantity determine its local electric field intensity that is subject to the screening by the local H$_2$O dipoles and modified by the solute-solute interactions [51].

Artifacts such as the cross section of mode reflectivity and the frequency dependence of transit polyaxiality contribute to the spectral intensity and the peak shape but not the transition of abundance, frequency and structure order induced by solvation [36]. Therefore, artifacts can be minimized by the peak area normalization [51].

## 2.2 O:H-O Bond Cooperative Relaxation

As a strongly correlated and fluctuating system, water prefers the statistic mean of the tetrahedrally-coordinated, two-phase structure in a bulk-skin or a core-shell manner of the same geometry but different O:H-O bond lengths[46-47]. It is essential to take water as a crystalline-like structure with well-defined lattice positions and inter- and intramolecular interactions with fluctuation. The O:H-O bond integrates the intermolecular weaker O:H nonbond (or called van der Waals bond with ~0.1 eV energy) and the intramolecular stronger H-O polar-covalent bond (~4.0 eV) with asymmetrical, short-range interactions and coupled by the Coulomb repulsion between electron pairs on adjacent oxygen ions [47], as Figure 1 illustrated.

The O:H nonbond and the H-O bond segmental disparity and the O-O coupling allow the segmented O:H-O bond to relax oppositely – an external stimulus dislocates both O ions in the same direction but by different amounts, see as Figure 1b. The softer O:H nonbond always relaxes more than the stiffer H-O bond with respect to the $H^+$ coordination origin. The ∠O:H-O containing angle θ relaxation contributes only to the geometry and mass density. The O:H-O bond bending has its specific vibration mode that does not interfere the H-O and the O:H stretching vibrations[47]. The O:H-O bond cooperativity determines the properties of water and ice under external stimulus such as molecular undercoordination[52-56], mechanical compression [28, 35, 57-59], thermal excitation [60-62], solvation [63-64] and determines the molecular behavior such as solute and water molecular thermal fluctuation, solute drift motion dynamics, or phonon relaxation.

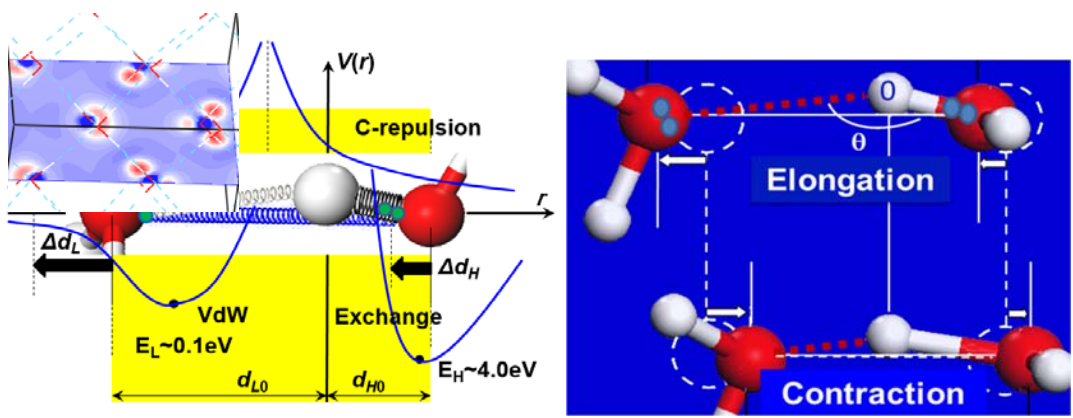

Figure 1. (a) Asymmetrical, short-range, coupled three-body potentials for the segmented O:H-O bond and (b) its segmental length cooperativity. Any relaxation of the O:H-O bond proceeds by elongating one part and contracting the other with respect to the $H^+$ coordination origin. The softer O:H always relaxes more than the stiffer H-O. The ∠O:H-O containing angle

θ does not contribute to the bond length and energy (reprinted with permission from ref 47).

## 2.3. Solvation Dynamics and Molecular Interaction

Solvation dissolves a substance into solutes of eitherly charged particles or dipolar molecules attached with H$^+$ and ":", distributed regularly in the yet seemingly disordered bulk liquid with or without skin preferential occupation. Charged particles serve each as source of electric field that aligns, clusters, stretches, and polarizes their neighboring solvent molecules to form the supersolid [43] or semirigid[40, 65] hydration shells. The polarized solvent molecules in the hydration shells screen in turn the solute electric field. Compared with solid surface chemisorption and programmed doping, no regular bonds such as covalent or ionic form between the solute and the solvent molecules but only form O:H-O and H↔H and O:⇔:O nonbonds with dominance of induction, repulsion, polarization, and hydrogen bond formation[36, 66].

It would be efficient to deal with solvation in the same way of handling chemisorption and defect formation by equaling the aqueous solutes to the adsorbates, dopants, point defects, and impurities in the solid phase, disregarding the structural fluctuation and drifting motion of the solute and solvent. In fact, what determines the properties of a solution is the intra- and intermolecular energies and charge polarization. Thermal fluctuation or solute drift motion are processes of energy dissipation without involvement of energy emission or absorption from the thermodynamic point of view. Another reason to treat the solvent water as crystal-like is its numbers of H$^+$ and ":" and the O:H-O configuration conservation unless excessive H$^+$ or ":" is introduced [47].

## 2.4. Ionic Polarization and Nonbond Repulsion

Recent progress[36-37, 42] shows that no charge sharing occurs or regular bond forms between the solute and the solvent molecules. Only solute-solvent induction and repulsion occur in the aqueous solutions. The following formulate the solvation dynamics of HBr, LiBr, and LiOH:

HBr + H$_2$O → Br$^-$ + H$_3$O$^+$ (H↔H anti-HB fragilization and Br$^-$ hydration-shell formation by polarization)
LiBr + H$_2$O → Br$^-$ + Li$^+$ + H$_2$O (Li$^+$ and Br$^-$ hydration-shell formation by ionic polarization)
LiOH + H$_2$O → Li$^+$ + HO$^-$ + H$_2$O (O:⇔:O super-HB compression and Li$^+$ hydration-shell formation).

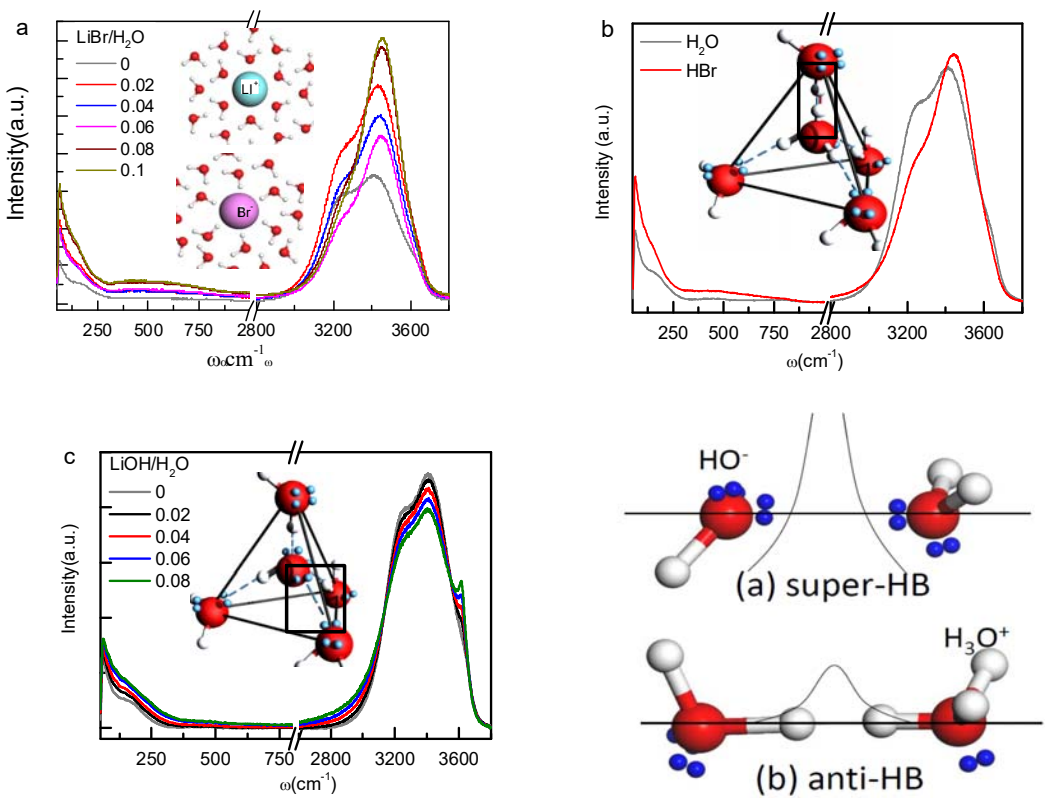

Figure 2. Full-frequency Raman spectroscopy for (a) LiBr, (b) HBr (0.06 molar fraction) [36], and (c) LiOH solutions [37] with insets showing the central substitution of the 2$H_2O$ with (b) $H_3O^+$ for acidic and (c) $HO^-$ for basic solutions, which derives (d) the O:⇔:O super-HB between the $OH^-$ and a $H_2O$ and the H↔H anti-HB between the $H_3O^+$ and a $H_2O$ molecule and their respective repulsive potentials. The inset illustrates (a) ionic polarization and hydration shell formation. Features below 200 $cm^{-1}$ and around 3200 $cm^{-1}$ characterize, respectively, the O:H nonbond and the H-O bond stretching vibrations.

For a specimen containing $N$ number of $H_2O$ molecules, there is a total of 2$N$ protons and 2$N$ lone pairs to form uniquely the O:H-O bonds throughout the specimen disregarding the phase structure of the water or ice. What one can change is the O:H-O segmental lengths and the ∠O:H-O angle [47]. Both $Br^-$ and $Li^+$ ions serve as each a point polarizer that aligns, stretches, and polarizes the surrounding hydrogen bonds to form their supersolid[43] or semirigid [40, 65] hydration shells, without changing the conservation but local structure distortion, as Figure 2a inset illustrated[42].

However, an introduction of an excessive ":" or $H^+$ braeks the 2$N$ number and the O:H-O configuration invariance. For instance, LiOH solvation adds a $HO^-$ with one $H^+$ and three ":", turning the 2N protons into 2N+1 and the 2N lone pairs into 2N+3, resulting in the 2N+3 – (2N+1) = 2 excessive lone pairs that can only form the O:⇔:O interaction without any other choice[37]. Likewise, HBr solvation creates the H↔H interaction[36]. The unprecedently H↔H interproton repulsion and O:⇔:O inter-lone-pair compression govern the performance of the acid and basic solutions[36-37].

Both the $H_3O^+$ and the $OH^-$ retain their $sp^3$-hybridized electron orbitals but have unbalanced numbers of protons and lone pairs, as the Figure 2 b and c insets illustrated[36-37]. The $H_3O^+$ and $OH^-$ substitution for the central $H_2O$ molecule in the $2H_2O$ unit cell creates regularly the (Figure 2 b inset) H↔H anti-HB point breaker and (Figure 2 c inset) the O:⇔:O super-HB point compressor. Figure 2d further illustrates the H↔H anti-HB and O:⇔:O interactions. No such conclusion could be possible if one assumed water as an amorphous substance or a randomly ordered system or deemed the $H^+$ or the ":" freely hopping or shuttling.

These point H↔H breakers, O:⇔:O compressors, and ionic polarizers govern the performance of the hydration network of acid, base, and salt solutions. The H↔H anti-HB disrupts the solution network and the surface stress [36], which is the same to the H-induced embrittlement of metals and alloys [67-68]. The O:⇔:O super-HB compresses the neighboring O:H-O bond [37] to have the same effect of mechanical compression that shortens the O:H nonbond and elongates the H-O bond [28].

3. **Results and Discussion**

3.1 O:H-O Phonon Abundance and Stiffness Transition

Figure 2 a-c displays the full-frequency Raman spectra for HBr, LiBr, and LiOH solutions. LiBr and HBr solvation stiffens the H-O phonon due to ionic polarization. LiOH solvation results in a new feature at 3610 cm$^{-1}$ in addition to the flattened main peak shifting downward. Insets show the (a) ionic hydration shell formation, (b) H↔H interproton interaction, (c) O:⇔:O inter-lone-pair interaction and (d) amplifies the H↔H and O:⇔:O repulsive nonbonds [36-37].

Figure 3 shows the LiBr, HBr[36] and LiOH[37] DPS profiles from which we obtain the fraction coefficients by the peak area integration. Ionic polarization transits cooperatively the O:H-O segmental phonons $\omega_H$ from 3200 to 3480 cm$^{-1}$ and $\omega_L$ from 200 to 100 cm$^{-1}$ in their hydration shells because of the O-O repulsion, as Figure 3 b inset illustrated. Ionic polarization has the same effect of molecular undercoordination to polarize and transit the O:H-O bond length and stiffness [47, 69]. Therefore, the ionic hydration shells behave identically to the supersolid water skin. The supersolid means highly ordered structure (longer $\omega_H$ lifetime [4]) of semirigid [40], high stress, polarized charge distribution, low density, slow molecular dynamics, and high thermal stability[42, 47].

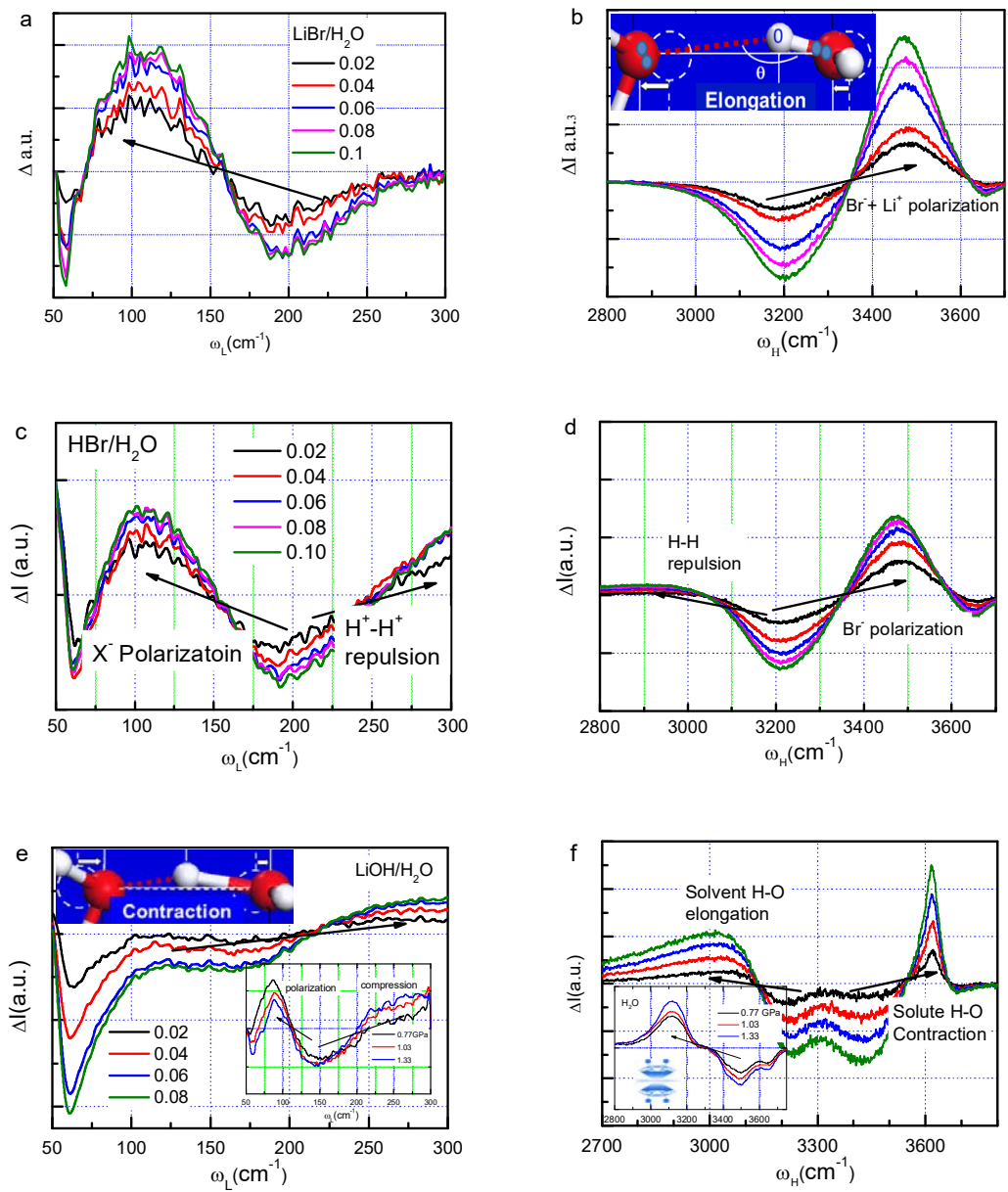

Figure 3. Concentration dependence of the $\omega_x$ DPS for (a, b) LiBr/$H_2O$, (c, d) HBr/$H_2O$[36], and (e, f) LiOH/$H_2O$[37] solutions. Insets b and e illustrate the manners of O:H-O bond elongation by ionic polarization and contraction by O:⇔:O compression, respectively. Inset DPS spectra in parts e and f result from mechanical compression at room temperature of liquid water.

The difference of the $\omega_H$ phonon abundance between the LiBr and the HBr solution in Figure 3b and d shows that the H-O phonon abundance of the HBr solution is less than the LiBr, which confirmed that the $H^+$ in the HBr solution does not polarize its neighboring $H_2O$ molecules because of the $H_3O^+$ formation. On the other hand, the H↔H repulsion shifts a tiny fraction, but O:⇔:O compression shifts considerable amount of the solvent H-O feature to 3100 cm$^{-1}$ and below. This high broadness indicates the long distance solvent O:H-O bond relaxing by the point compression. The <3100 cm$^{-1}$ phonon abundance difference between Figure 3d and f discriminates the strength of the H↔H and the O:⇔:O repulsive interactions. The latter is estimated four time of the former by considering the charge quantities of the same separation. The DPS for LiOH solution in Figure 3f also shows an excessive sharp peak at 3610 cm$^{-1}$, which indicates the rather local nature of the solute H-O bond contraction. The spectral shift annihilates the effect of $Li^+$ polarization. Excitingly, the O:⇔:O compression is much greater than the critical pressure, 1.33 GPa, for room-temperature water-ice transition. As shown in Figure 3 e and f insets mechanical compression transits the H-O phonon from 3300 cm$^{-1}$ to below [28].

The two DPS peaks clarify that the longer 200 ± 50 fs lifetime features the slower molecular motion but higher-frequency 3610 cm$^{-1}$ solute H-O bond vibration and the other shorter time on 1–2 ps scales is related to the lower-frequency <3100 cm$^{-1}$ elongated solvent H-O bond vibration upon HO$^-$ solvation[37] in NaOH solutions [32-33].

### 3.2. Solute-Solvent and Solute-Solute Interactions

Figure 4 compares the concentration dependent $f_{LiBr}(C) = f_{Li}(C) + f_{Br}(C)$, $f_{HBr}(C) = f_{Br}(C)$ and $f_{LiOH}$(<3100 cm$^{-1}$, 3610 cm$^{-1}$) that feature the relative number of O:H-O bonds transiting from the ordinary water into the hydrating states. The $f_x(C)$ concentration trends recommend the following, see Figure 4:

1) The $f_H(C) \equiv 0$ means that the $H^+(H_3O^+)$ is incapable of polarizing its neighboring HBs but only breaking and slightly repulsing its neighbors[36].
2) The $f_{Li}(C) \propto C$ means the constant shell size of the small $Li^+$ cation (radius = 0.78 Å) without being interfered with by other solutes. The constant slope indicates that the number of bonds per solute is conserved in the hydration shell. The electric field of a small $Li^+$ cation is fully screened by the $H_2O$ dipoles in its hydration shells; thus, no cation-anion or cation-cation interaction is involved for the LiBr and HBr solutions.
3) The $f_{OH}(C) \propto C$ (<3100, 3610 cm$^{-1}$) means that the numbers of the O:⇔:O compression-elongated solvent H-O bonds ($f_{OH}(C) = 0.985C$) and the bond-order-deficiency shortened solute H-O bonds ($f_{OH}(C) = 0.322C$) are proportional to the solute concentration. Bond order deficiency shortens and stiffens the bonds between undercoordinated atoms [47].

4) The $f_{Br}(C) \propto 1-\exp(-C/C_0)$ toward saturation means the number of H₂O molecules in the hydration shells is insufficient to fully screen the Br⁻ (radius = 1.96 Å) solute local electric field because of the geometric limitation to molecules packed in the crystal-like water. This number inadequacy may further evidence the well-ordered crystal-like solvent. The solute can thus interact with their alike – only anion-anion repulsion exists in the Br⁻ - based solutions to weaken the local electric field of Br⁻. Therefore, the $f_{Br}(C)$ increases approaching saturation, the hydration shells size turns to be smaller, which limits the solute capability of bond transition.

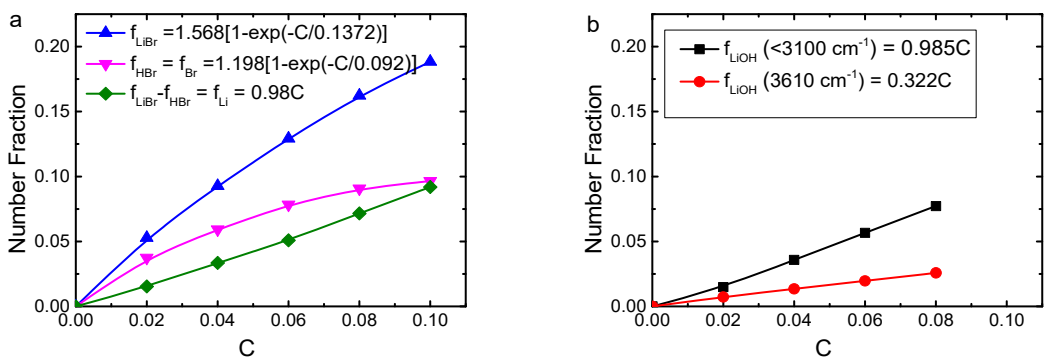

Figure 4. Concentration dependence of the fraction coefficients for LiBr/H₂O (3500 cm⁻¹), HBr/H₂O (3500 cm⁻¹), and LiOH (<3100, 3610 cm⁻¹) solutions. The liner $f_x(C)$ indicates the invariance of the Li⁺ and OH⁻ hydration shell size and the exponential $f_x(C)$ features Br⁻-water interaction with contribution of Br⁻ - Br⁻ interaction.

Therefore, the $f_x(C)$ and its slope give profound information not only on the solute-solute and solute-solvent interaction but also on the relative number of bonds transiting from the referential mode of water to the hydration, by ionic polarization or O:⇔:O compression.

### 3.3. Surface Stress, Solution Viscosity, Molecular Diffusivity

Figure 5 a compares the concentration dependence of the contact angle between the solutions and glass substrate measured at 298 K. The surface stress is proportional to the contact angle. One can ignore the reaction between the glass surface and the solution, as we want to know the concentration trends of the stress change at the air-solution interface of a specific solution. Ionic polarization and O:⇔:O compression enhance the stress, but the H↔H point fragilization destructs the stress as the Li⁺ hydration forms an independent hydrating fragment. The H↔H fragmentation has the same effect of thermal fluctuation on depressing the surface stress[29] with different mechanisms. Thermal excitation weakens the individual O:H bond throughout the bulk water, but H↔H fragilization weakens the O:H bonds between the Li⁺ hydrated fragments. Ionic and O:⇔:O polarization has the same effect of molecular undercoordination on constructing the surface stress. Both polarization and undercoordination form the supersolid phase; the former occurs in the hydration shell throughout the bulk, but the latter only takes place in skins. Ions may prefer occupying the skin of the solution, which is a different situation.

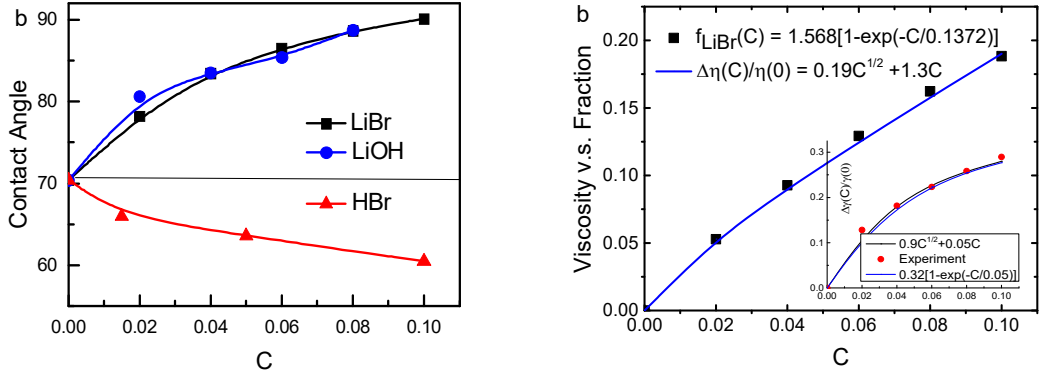

Figure 5. Concentration dependence of (a) solution contact angles on glass substrate and (b) trend agreement between the relative viscosity of Jones–Dole notion[70] and the $f_{LiBr}(C)$ contact angle for the LiBr/H$_2$O solutions. Inset b shows that the LiBr/H$_2$O surface stress follows the same exponential trend of the $f_{LiBr}(C)$ and the relative viscosity with different coefficients because of the additional molecular undercoordination effect that enhances the ionic polarization.

In aqueous solutions, solute molecules are taken as Brownian particles drifting randomly under thermal fluctuation by collision of the solvent molecules. The viscosity of salt solutions is one of the important macroscopic parameters often used to classify water-soluble salts into structure making or structure breaking. The drift motion diffusivity D($\eta$, R, T) and the solute-concentration-resolved solution viscosity $\eta$(C) follow the Stokes-Einstein relation [44] and the Jones–Dole expression[70], respectively,

$$\begin{cases} \dfrac{D(\eta,R,T)}{D_0} = \dfrac{k_B T}{6\pi\eta R} & (Drift) \\ \dfrac{\Delta\eta(C)}{\eta(0)} = A\sqrt{C} + BC & (Viscosity) \end{cases}$$

where $\eta$, R, and $k_B$ are the viscosity, solute size, and Boltzmann constant, respectively. $D_0$ is the coefficient in pure water. The coefficient A and its nonlinear term is related to the solute mobility and solute-solute interaction. The coefficient B and the linear term reflects the solute-solvent molecular interactions. The $\eta(0)$ is the viscosity of neat water.

SFG measurements [71-72] revealed that the SCN⁻ and CO$_2$ solution viscosity increases with solute concentration or solution cooling. The H-O phonon relaxation time increases with the viscosity, and results in molecular motion dynamics. Therefore, ionic polarization stiffens the H-O phonon and slows down the molecular motion in the semirigid or supersolid structures.

One may note that the relative viscosity and the measured surface stress due to salt solvation are in the same manner of the $f_{LiBr}(C)$, $f_{LiBr}(C) = a\left[1 - \exp(-C/C_0)\right]$. One can adjust the Jones–Dole viscosity coefficients $A$ and $B$ and fit the surface stress to match the measured $f_{LiBr}(C)$ curve in Figure 5 b. The trend consistency clarifies that the linear term corresponds to Li⁺ hydration shell size and the nonlinear part to the resultant of Br⁻-water and Br⁻-Br⁻ interactions. It is clear now that both the solution viscosity and the surface stress are proportional to the extent of polarization or to the sum of O:H-O bonds in the hydration shells. Therefore, polarization raises the surface stress, solution viscosity and rigidity, H-O phonon frequency, and H-O phonon lifetime but decreases the molecular drift mobility, consistently by shortening the H-O bond and lengthening the O:H nonbond.

### 3.4. LiOH Solvation Heating up Its Solution

According to chemical bond theory,[73] energy stores in the chemical bonds and the energy emission or absorption proceeds by bond relaxation – the equilibrium atomic distance and binding energy change [68]. Bond dissociation and bond elongation release energy but bond formation and bond contraction absorb energy, leading to the exothermic and endothermic reaction. Molecular diffusive motion or structure fluctuation only dissipate energy with negligible energy absorption or energy emission.

LiOH solvation undergoes the exo- and endo-thermic reactions besides thermal dissipation by structural fluctuation and molecular diffusion and non-adiabatic calorimetric detection. The endothermic processes include ($Q_{a,i}$):

i) The hydrating H-O bond contraction by Li⁺ polarization,
ii) Solute H-O contraction by bond-order-deficiency,
iii) Solvent H-O thermal contraction by temperature increases.

The exothermic processes include ($Q_{e,j}$):

i) LiOH dissolution into Li⁺ and OH⁻
ii) Solvent H-O elongation by O:⇔:O compression,
iii) O:H elongation by Li⁺ polarization and thermal excitation.

The total energy should conserve:

$$\sum_3 Q_{e,i}(C) - \sum_3 Q_{a,j}(C) - \sum_2 Q_{dis,l}(C) = 0$$

These exo- and endothermic processes shall compensate each other to a certain extent during solvation. We neglect the energy dissipation and the thermodynamics of Li$^+$ solvation, as LiBr dissociation and ionic polarization derive no apparent temperature change. One can focus on the exothermic solvent H-O elongation by O:⇔:O compression and solute H-O contraction by bond-order-deficiency and the temperature change and thus estimate the energy emitted by the H-O elongation.

To seek for the correspondence between the solution temperature T(C) and the f$_{LiOH}$(C) solutions, we conducted the *in-situ* solution calorimetric detection using a regular thermometer to monitor the solution temperature in a glass beaker under the ambient temperature of 25 °C. The solution was stirred using a magnetic bar rotating in the beaker in 5 Hz frequency. Figure 6 plots the LiOH solution temperature T(C) with comparison of the bond transition fraction coefficients f(C) for the elongated solvent H-O bonds and for the solute H-O bond contraction upon LiOH solvation.

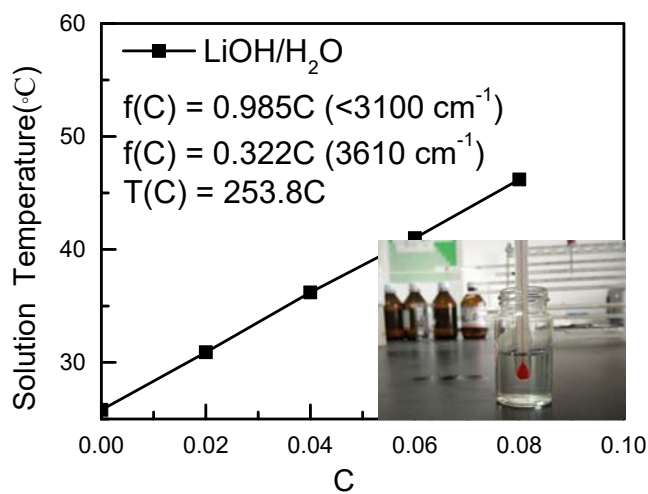

Figure 6. Linear correspondence between the highest solution temperatures T(C) and the concentration of the LiOH/H$_2$O solutions. Noted are the linear dependence of the fraction coefficient f(C) for the H-O phonon bands at <3100 cm$^{-1}$ and 3610 cm$^{-1}$ waveneumbers.

With the measured T(C) = 253.8C, f$_e$(C) = 0.985C (<3100 cm$^{-1}$), and f$_a$(C) = 0.322C (3610 cm$^{-1}$), one can estimate the energy emission from the solvent O:H-O bond elongated by O:⇔:O compression by ignoring the thermal fluctuation featured at 200 cm$^{-1}$. LiOH solvation transits the 200 cm$^{-1}$ mode for the O:H stretching (~0.095 eV[47]) partially to 110 and 300 cm$^{-1}$ which absorbs/emits a negligible amount of O:H energy.

The energy cost $Q_0$ to heat up a unit mass of the solution (m = 1) from $T_i$ to $T_f$ by increasing the solute contraction up to $C_M$ equals ($h_0 = 4.18$ J(gK)$^{-1}$ = 0.00039 eV(bondK)$^{-1}$ is the specific heat for liquid water):

$$\int_0^{Q_0} dq = h_0 \int_0^1 dm \int_0^{C_M} \frac{dt(C)}{dC} dC = h_0 T(C_M)$$

The energy difference between exothermic solvent H-O elongation $Q_e$ and the endothermic solute H-O contraction $Q_a$ heats up the solution ($h_e$ and $h_a$ are the energy emission and absorption per bond),

$$\int_0^{Q_e} dq_e - \int_0^{Q_a} dq_a = \left[ h_e \int_0^{f_e} dm_e - h_a \int_0^{f_a} dm_a \right] \int_{T_i}^{T_f} dt = [h_e f_e - h_a f_a] T(C_M)$$

Equaling the energy emission by H-O elongation to its loss by H-O contraction and heating solution with an approximation of $h_a \approx h_e$, yields,

$$h_e = \frac{h_0}{f_e - f_a h_a / h_e} \approx \frac{h_0}{(0.985 - 0.322) C_M} = 19 h_0$$

The energy emitted by an elongated H-O bond,

$$q_e = h_e T(C_M) = 253.8 C_M \times 19 h_0 = 0.151 (eV/bond),$$

which is 158% of the O:H cohesive energy 0.095 eV at room temperature.[47] It is thus verified that the energy remnant of the solvent H-O exothermic elongation and the solute H-O endothermic contraction heats up the solution.

We can also estimate the energy emission from the H-O bond elongation with the documented values of ($d_H$, $E_H$, $\omega_H$) = (1.0 Å, 4.0 eV, 3200 cm$^{-1}$) for the bulk water [74], and for basic hydration (1.05 Å, $E_2$, 2500/3000 cm$^{-1}$) using the frequency function, $\omega^2 \propto E/d^2$:

$$\Delta E = 2E\left(\frac{\Delta d}{d} + \frac{\Delta \omega}{\omega}\right) = 8\left(\frac{0.05}{1} - \frac{(700; 200)}{3200}\right) = -(1.35;\ 0.35)\ eV/bond$$

(3)

The H-O bond elongation losses its cohesive energy from 4.0 by 0.35~1.35 eV. Although the 0.15 eV energy emission may be underestimated, the O:⇔:O compression elongated H-O elongation is certainly the intrinsic and dominant source for heating up the LiBr solution.

### 3.5. Prospectus

Before conclusion, we would like to recommend the urgency of amplifying the conception and concurrently employed approaches for solvation study. One would benefit more from the perspective of molecular motion and structure fluctuation to the solvation bonding dynamics and thermodynamics. The latter ensures comprehensive information on the cooperativity of intermolecular nonbonding and intramolecular bonding interactions and the associated polarization or depolarization. Intramolecular covalent bond formation and contraction absorb energy, but the covalent bond dissociation and elongation emit energy. Molecular vibration, drift motion, reorientation, and even evaporation dissipate energy caped at the O:H cohesive energy of 0.1 eV in the solution.

Classical thermodynamics in terms of Gibbs free energy or total energy minimization depends functionally on the applied stimuli of pressure, temperature, composition, etc.; inclusion of individual bond relaxation and inner energy transportation, particularly, the repulsive and the weak intermolecular interactions would be necessary. Molecular dynamics calculations treat the molecules as independent and invariant motifs; inclusion of the intramolecular covalent bonding interaction and the cooperativity with intermolecular motion, would be a helpful practice. Expanding the density function theory to embrace the temperature change and strongly localized anisotropic inter- and intramolecular interactions would be profoundly revealing.

Similar to molecular dynamics, the time-resolved IR spectroscopy deals with molecular motion and intermolecular interaction through monitoring the decay/life/relaxation time of the intramolecular H-O vibrational spectral signal upon switching off/on the excitation/detection, which is in the similar principle of optical fluorescent spectroscopy. Both the relaxation time of an optical fluorescent spectrum and the phonon life relaxation share the same mechanism of operation. The optical relaxation time depends on the density and distribution of defects and impurities in a solid specimen as the defects inhibit the electron transition from the excited states to the ground for exciton recombination[45]. The phonon relaxation time varies with the viscosity of the solution. Cooling the solution or increasing the salt concentration raises the solution viscosity and elongates the phonon lifetime of relaxation[71-72], which lowers the Stokes-Einstein drift diffusivity.[44] Extending the time-dependent IR spectroscopy to detect the intramolecular bond relaxation[4] would lead to its bright capability.

A combination of the ultrafast 2GIR spectroscopy and the DPS strategy will ensure comprehensive information. The former probes the molecular spatial and temporal motion kinetics, and the latter resolves the on-site O:H-O bond length and energy dynamics. A further extension of the developed knowledge and strategies to the solvation bonding and nonbonding dynamics of complicated solutes, solution-protein interactions, drug-cell targeting, and biomolecular cell activating and inactivating molecular interactions toward discoveries in molecular crystals and liquids would certainly be even more fascinating, revealing, and rewarding.

4. **Conclusion**

We have thus resolved quantitatively the number fraction and phonon stiffness of ordinary HBs transiting into the hydration shells of HBr, LiBr, and LiOH solutions. The relaxation of the O:H-O bonds in the network caused by $H^+(H_3O^+)$, $OH^-$, $Li^+$, and $Br^-$ solutes can thus be discriminated as follows:

1) $H_3O^+$ hydronium formation in acid solution creates the H↔H anti-HB that serves as a point breaker to disrupt the HBr solution network and the surface stress. The $Br^-$ polarization dictates the O:H (from 200 partially to 110 and 300 cm$^{-1}$) and the H-O phonon (from 3200 to 3480 cm$^{-1}$) frequency cooperative shift.
2) $OH^-$ hydroxide forms the O:⇔:O super-HB point compressor to soften the nearest solvent H-O bond (from above 3100 cm$^{-1}$ to below), and meanwhile, the solute H-O bond shortens to its dangling radicals featured at 3100 cm$^{-1}$. The $Li^+$ polarization effect has been annihilated by the O:⇔:O compression and the solute H-O contraction.
3) $Li^+$ and $Br^-$ serve each as a point polarizer that aligns, stretches, and polarizes the surrounding O:H-O bonds and makes the hydration shell supersolid. The polarization transits the $\omega_L$ from 200 to 100 cm$^{-1}$ and the $\omega_H$ from 3200 to 3480 cm$^{-1}$.
4) The solute capability of bond transition follows: $f_H(C) = 0$, $f_{Li}(C) \propto f_{OH}(C) \propto C$, and $f_{Br}(C) \propto 1-\exp(-C/C_0)$ toward saturation. The size trends of the f(C) reveal the hydration shell size and the solute-solvent, and solute-solute interactions.
5) The concentration trends consistent among the salt solution viscosity, surface stress, and the $f_{LiBr}(C)$ suggest their common origin of polarization associated with O:H-O bond transition from water to hydration shells.

6) The energy remnant of the solvent H-O exothermic elongation by O:⇔:O repulsion and the solute H-O endothermic contraction by bond-order-deficiency heats up the solution, which has little to do with the solute or solvent molecular motion dynamics.

We have thus verified the essentiality of the super-HB and anti-HB dictating solute-solvent molecular interactions and their capabilities of transforming the HBs and the surface stress of these aqueous solutions systematically. The reported fine resolution and the consistent insight are only possible by taking the solutions as highly ordered, strongly correlated, and fluctuating systems with consideration of the O:H-O bond cooperativity. The solutes serve each as an impurity with their electric fields of polarization or repulsion and the local screening by the hydrating $H_2O$ dipoles. Observation may extend to general understanding of the solvation process of acids, bases, and salts according to their abilities of creating the excessive $H^+$ protons, lone pairs, and ionic polarizers when they are hydrated. The nonbonding fragilization, compression, and polarization shall be critical to the hydration networks that functionalize DNA, proteins, cells, drugs, ionic channels, etc. The DPS forms such a powerful yet straightforward means to remove the artifacts and resolve the fraction and stiffness of the ordinary HBs transformation and their impact on solution properties.

## Acknowledgement

Financial support was received from Natural Science Foundation (No. 11502223), the Science Challenge Project (No. TZ2016001) of China, Zhejiang Province (No. LY18E060005), Hunan Province (No. 2016JJ3119), and Shen Zhen (No. 827000131).